\documentclass[lettersize,journal]{IEEEtranV1.7a}
\usepackage{amsmath,amsfonts,amssymb,amsthm}
\usepackage{array}
\usepackage{url}
\usepackage{graphicx}
\usepackage{color,colortbl,xcolor}
\usepackage{cite}

\usepackage{tikz}

\def\thline{\noalign{\hrule height 1pt}}

\def\a{\alpha}
\def\b{\beta}
\def\Cb{C_\mathrm{b}}
\def\Cr{C_\mathrm{r}}
\def\Dc{D_C}
\def\Du{D_U}
\def\Dv{D_V}

\newcommand{\tn}[1]{\textcolor{black}{#1}}
\newcommand{\tb}[1]{\textcolor{blue}{#1}}
\newcommand{\TB}[1]{\mathbf{\textcolor{blue}{#1}}}
\newcommand{\tr}[1]{\textcolor{red}{#1}}

\begin{document}

\title{
Spatially-Aligned Chroma from Luma Prediction for\\Lossless JPEG~XS Raw Image Compression
}

\author{
Taizo~Suzuki,~\IEEEmembership{Senior Member,~IEEE},
Soma~Yokota,
and
Masaki~Onuki
\thanks{T. Suzuki and S. Yokota are with the Institute of Systems and Information Engineering, University of Tsukuba, Ibaraki 305-8573, Japan (e-mail: \{taizo,yokota\}@cs.tsukuba.ac.jp).}
\thanks{M. Onuki is with SonicAI Inc., Tokyo 108-0073, Japan (e-mail: m-onuki@sonicai.jp).}
}

\markboth{IEEE Signal Processing Letters,~Vol.~XX, No.~XX, April~2026}%
{How to Use the IEEEtran \LaTeX \ Templates}

\maketitle

\begin{tikzpicture}[remember picture,overlay]
\node[
  anchor=south west,
  text width=0.92\paperwidth,
  align=left,
  font=\scriptsize
] at ([xshift=0.55in,yshift=0.15in]current page.south west)
{\copyright{} 2026 IEEE. Personal use of this material is permitted.
Permission from IEEE must be obtained for all other uses, in any current or future media,
including reprinting/republishing this material for advertising or promotional purposes,
creating new collective works, for resale or redistribution to servers or lists, or reuse
of any copyrighted component of this work in other works.};
\end{tikzpicture}

\begin{abstract}
This study proposes a Chroma from Luma (CfL)-enhanced Star-Tetrix transform (STT), referred to as CfL-STT, for improving raw image compression in JPEG~XS.
The proposed CfL-STT integrates CfL prediction into the STT to predict chroma components from the luma component in CFA-sampled raw images.
Unlike conventional CfL prediction designed for full-color images, the proposed method employs spatially-aligned luma samples obtained via linear interpolation along the horizontal and vertical directions to match the chroma sampling grid.
This spatial alignment suppresses high-frequency sensor noise while preserving cross-channel correlation, resulting in a more decorrelated Y$\Delta$DuDv color space.
The proposed method was implemented in the JPEG~XS reference software and evaluated on raw image datasets.
Experimental results demonstrate that a direct application of CfL prediction yields image-dependent performance and may degrade coding efficiency due to the lack of spatial alignment, whereas the proposed CfL-STT consistently improves coding efficiency in lossless raw image compression while preserving exact reversibility.
\end{abstract}


\section{Introduction}
\begin{figure}[t!]
		\centering
        \includegraphics[scale=0.275,keepaspectratio=true]{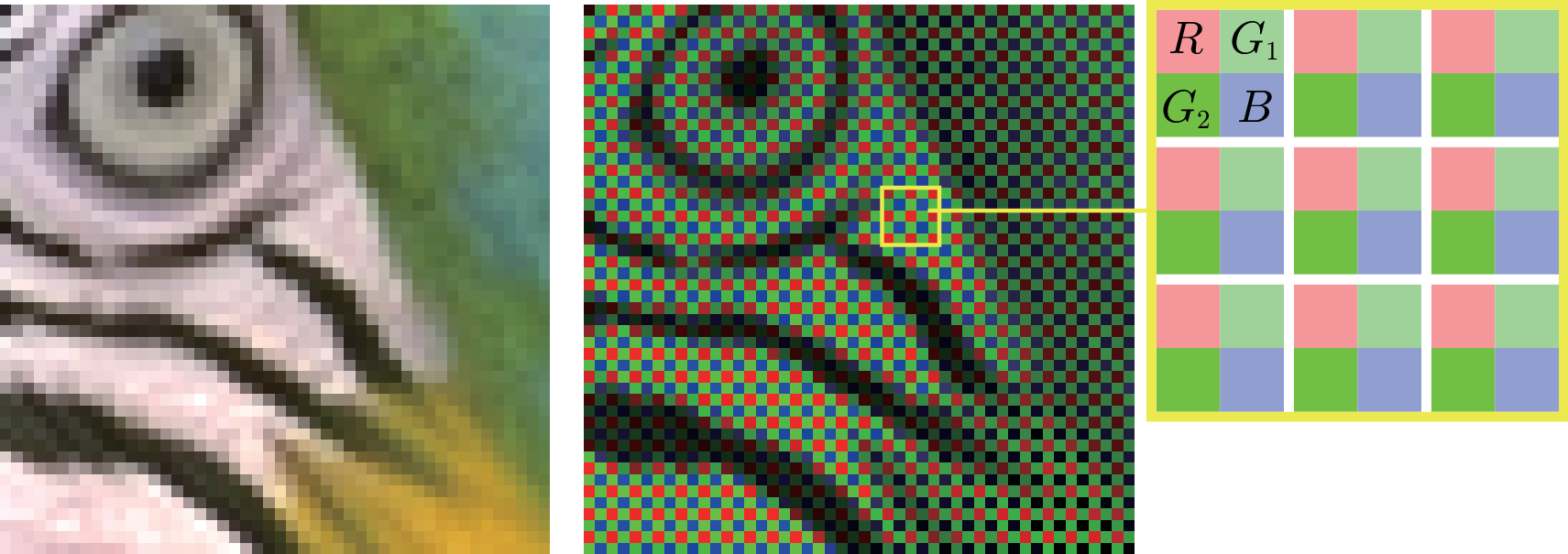}
		\caption{Bayer pattern of a region from \textit{Parrot} in the Kodak images dataset~\cite{KODAKTestImages} (each $2\times 2$ pixel block corresponds to a macropixel): (left) RGB full-color image and (right) simulated raw image with corresponding diagram.}
		\label{Fig_CFA}
	\end{figure}

\IEEEPARstart{F}{ULL-COLOR} (RGB) images commonly used today are obtained by applying a simple yet inherently \textit{lossy} camera processing pipeline, including black level correction, white balance, demosaicing, and gamma correction, to raw images sampled by a color filter array (CFA), which provides only one red, green, or blue component per pixel.
However, with the growing demand for flexible post-editing, there is increasing interest in both lossless and lossy compression of CFA-sampled raw images (hereinafter referred to as raw images).
Consequently, compression-first approaches are gaining importance alongside demosaicing-first approaches such as JPEG~\cite{Wallace1992IEEE}, JPEG 2000~\cite{Skodras2001IEEE}, and JPEG AI~\cite{Ascenso2023MM}.
This trend has been recognized in JPEG XS 2nd Edition, which introduced compression-first raw image coding as an optional operating mode~\cite{Descampe2021ProcIEEE}.
In this study, we focus on the Bayer CFA, the most widely adopted RGGB layout (Fig.~\ref{Fig_CFA}).

Understanding spectral–spatial transforms is essential for effective raw image compression~\cite{Zhang2006TIP,Malvar2012DCC,HernandezCabronero2018TMM,Lee2018TIP,Lee2020SPL,Richter2021TIP,Suzuki2020TIP,Suzuki2022TIP}.
Such transforms map CFA-sampled RGGB data into a decorrelated Y$\Delta$UV representation, typically expressed as Y$\Delta$CbCr or Y$\Delta$CoCg, consisting of luma, difference luma (green), and two chroma components.
Among them, the Star-Tetrix transform (STT)~\cite{Richter2021TIP}, built on three lifting-based 5/3-tap wavelet transforms, has been adopted in JPEG~XS 2nd Edition for raw image compression due to its strong decorrelation capability and low computational cost.
In parallel, the AOMedia Video 1 (AV1)~\cite{Han2021ProcIEEE} and the JPEG XL~\cite{Sneyers2025arXiv} introduced Chroma from Luma (CfL) prediction~\cite{Trudeau2018DCC}, which linearly predicts chroma components from luma to further reduce spectral redundancy in full-color images.
Related cross-component prediction tools have also been studied in the context of H.266/VVC, including subset-based cross-component linear model prediction~\cite{Huo2022TII} and adaptive chroma prediction based on luma differences~\cite{Huo2023TIP}.
However, raw images exhibit structured sampling patterns and nonuniform sensor noise, which violate the assumptions of conventional CfL prediction and degrade its prediction accuracy when directly applied in the raw domain.
Moreover, although recent learned image codecs and large language model-based predictors~\cite{Deletang2024ICLR} achieve impressive rate–distortion performance at low bitrates, they primarily target semantic compression, whereas raw image coding requires faithful signal reconstruction with low complexity and strict reversibility.

This study proposes a CfL-enhanced STT (CfL-STT), in which CfL prediction is integrated into the STT to predict chroma components from the luma component.
Unlike conventional CfL prediction, which assumes co-located color components, the proposed method employs spatially-aligned luma samples obtained via linear interpolation along the horizontal and vertical directions to match the chroma sampling grid.
This spatial alignment not only improves prediction robustness in CFA-sampled raw images but also introduces an implicit low-pass filtering effect that suppresses high-frequency sensor noise while preserving cross-channel correlation.
As a result, the Y$\Delta$UV representation produced by STT is transformed into a more decorrelated Y$\Delta$DuDv color space, where Du and Dv denote chroma residuals after spatially-aligned CfL prediction.
Since the proposed method relies only on interpolation and simple linear regression, it incurs limited additional computational complexity compared with the baseline STT.
Experimental results using the JPEG~XS reference software demonstrate that the proposed CfL-STT consistently improves compression performance in lossless raw image coding.

\section{Review and Preliminaries}
\subsection{Star-Tetrix Transform (STT)}\label{Sec_STT}
	\begin{figure*}[t]
		\centering
        \includegraphics[scale=0.3,keepaspectratio=true]{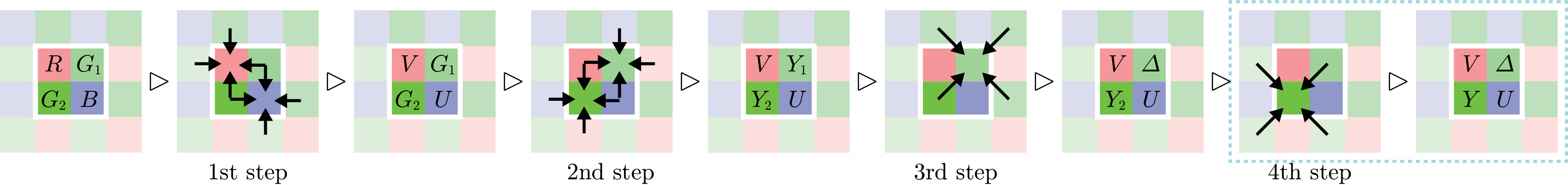}
		\caption{Implementation of STT (arrows indicate lifting steps in the STT process).}
		\label{Fig_Implements_STT}
	\end{figure*}

The Star-Tetrix transform (STT)~\cite{Richter2021TIP} is a fully reversible multi-component lifting transform designed to decorrelate both spectral and spatial redundancies inherent in Bayer CFA data.
It can be regarded as a CFA-domain extension of the JPEG 2000 reversible color transform (RCT)~\cite{Suzuki2022TIP}, where bilinear interpolation and spatial prediction are incorporated into a unified lifting structure.

Figure~\ref{Fig_Implements_STT} illustrates the implementation of STT.
Let $R$, $G$, and $B$ denote CFA samples at red, green, and blue sensor positions, and let the superscripts $l$, $r$, $t$, and $b$ indicate the neighboring samples to the left, right, top, and bottom, respectively.
At red and blue positions, the green signal is bilinearly interpolated from its four neighboring samples and subtracted from the measured value to form chroma components\footnote{In \cite{Richter2021TIP}, the two chroma components are denoted by $\Cb$ and $\Cr$, which correspond to $U$ and $V$ in this study.}:
    \begin{align}
        U
        &=
        B
        -
        \left\lfloor
            \frac{G_2^l + G_2^r + G_1^t + G_1^b}{4}
        \right\rfloor,
        \label{Eqs_STT_Cb}
        \\
        V
        &=
        R
        -
        \left\lfloor
            \frac{G_1^l + G_1^r + G_2^t + G_2^b}{4}
        \right\rfloor,
        \label{Eqs_STT_Cr}
    \end{align}
where $\lfloor\cdot\rfloor$ denotes a floor function.
This step corresponds to the first lifting stage of a YUV-like transform that is adapted to the CFA sampling structure.
By forming differences between the measured red and blue samples and the bilinearly interpolated green signal, the strong inter-channel correlation between color components is effectively removed.

Each green sample is then updated using a weighted average of the surrounding chroma components.
Let $w_\mathrm{r}$ and $w_\mathrm{b}$ be non-negative integer white-balance exponents that represent powers of two derived from camera metadata.
Two luma components are generated as
    \begin{align}
        Y_1
        &=
        G_1
        +
        \left\lfloor
            \frac{2^{w_\mathrm{r}}(V^l + V^r) + 2^{w_\mathrm{b}}(U^t + U^b)}{8}
        \right\rfloor,
        \label{Eqs_STT_Y1}
        \\
        Y_2
        &=
        G_2
        +
        \left\lfloor
            \frac{2^{w_\mathrm{r}}(V^t + V^b) + 2^{w_\mathrm{b}}(U^l + U^r)}{8}
        \right\rfloor.
        \label{Eqs_STT_Y2}
    \end{align}
These equations perform color decorrelation and white-balance adaptation simultaneously.
The powers-of-two weights $2^{w_\mathrm{r}}$ and $2^{w_\mathrm{b}}$ shift the effective luma axis toward the camera white point, which improves decorrelation between luma and chroma components.
This study sets $w_\mathrm{r}=w_\mathrm{b}=0$ for simplicity.

The next lifting stage introduces spatial decorrelation between the two luma fields.
A local prediction of $Y_1$ is formed from the surrounding $Y_2$ samples:
    \begin{equation}
        \varDelta = Y_1 - \left\lfloor\frac{Y_2^{l,t} + Y_2^{r,t} + Y_2^{l,b} + Y_2^{r,b}}{4}\right\rfloor.
        \label{Eqs_STT_Delta}
    \end{equation}
This term represents a high-frequency spatial luma difference that is analogous to a diagonal difference signal.
It captures the mismatch between the two interleaved luma grids and is therefore well suited for efficient entropy coding.

Finally, the remaining luma field is updated using the neighboring $\varDelta$ values as
    \begin{equation}
        Y = Y_2 + \left\lfloor\frac{\varDelta^{l,t} + \varDelta^{r,t} + \varDelta^{l,b} + \varDelta^{r,b}}{8}\right\rfloor.
    \end{equation}

After these four lifting steps, the CFA image is mapped into four decorrelated components $Y$, $\varDelta$, $U$, and $V$.
$Y$ represents low-frequency luma component, $U$ and $V$ represent chroma components, and $\varDelta$ represents a spatial luma difference that already exhibits high-pass characteristics.
Importantly, all steps are defined in an integer lifting framework, ensuring exact reversibility, which is essential for lossless raw image compression.

\subsection{Chroma from Luma (CfL) Prediction}\label{Sec_CfL}
Chroma from Luma (CfL) prediction~\cite{Trudeau2018DCC} is a linear regression model that predicts chroma components from the luma component in full-color image compression, thereby reducing residual spectral correlation.
In this framework, a chroma component $C$, which can be either $U$ or $V$, is treated as the dependent variable and is linearly predicted from the luma component $Y$, which serves as the independent variable.
The linear prediction model is given by
    \begin{equation}
        \widetilde{C} = \a_C~Y + \b_C,
        \label{Eqs_SLR}
    \end{equation}
where $\widetilde{C}$ denotes the predicted chroma component and $\a_C$ and $\b_C$ are the regression parameters.
The parameters are estimated by minimizing the chroma prediction error in the least-squares sense.
The corresponding cost function is defined as
    \begin{equation}
        f(\a_C,\b_C)
        =
        \sum_{n}
        \left|
            C(n) - \widetilde{C}(n)
        \right|^2
        ,
        \label{Eqs_EstErrNorm2}
    \end{equation}
where $n$ is the sample index.
The optimal parameters are uniquely obtained as
    \begin{align}
        \a_C
        &=
        \frac{
            N \sum_{n}\left(Y(n) C(n)\right)
            -
            \sum_{n}Y(n) \sum_{n}C(n)
        }{
            N \sum_{n}Y(n)^2
            -
            \left(\sum_{n}Y(n)\right)^2
        },
        \label{Eqs_Alpha_C}
        \\
        \b_C
        &=
        \frac{\sum_{n}C(n) - \a_C \sum_{n}Y(n)}{N},
        \label{Eqs_Beta_C}
    \end{align}
where $N$ is the total number of samples.
After parameter estimation, the chroma residuals
    \begin{equation}
        \Dc = C - \lfloor\widetilde{C}\rfloor
    \end{equation}
are encoded instead of the original chroma component $C$.
Note that to ensure integer reversibility, the predicted chroma component $\widetilde{C}$ is floored prior to subtraction.
The regression parameters $\a_C$ and $\b_C$ are transmitted as side information and allow exact reconstruction of the chroma signal from the luma and the decoded residual.

This formulation assumes co-located luma and chroma samples and spatially uniform noise characteristics, assumptions that do not hold in CFA-sampled raw images.
In particular, the spatial misalignment between luma and chroma samples and the presence of signal-dependent sensor noise can significantly degrade the accuracy and robustness of the regression model.
Therefore, directly applying CfL prediction to CFA-domain signals may lead to suboptimal decorrelation performance.

\section{CfL-Enhanced STT (CfL-STT)}
    \begin{figure}[t]
		\centering
	      \includegraphics[scale=0.3,keepaspectratio=true]{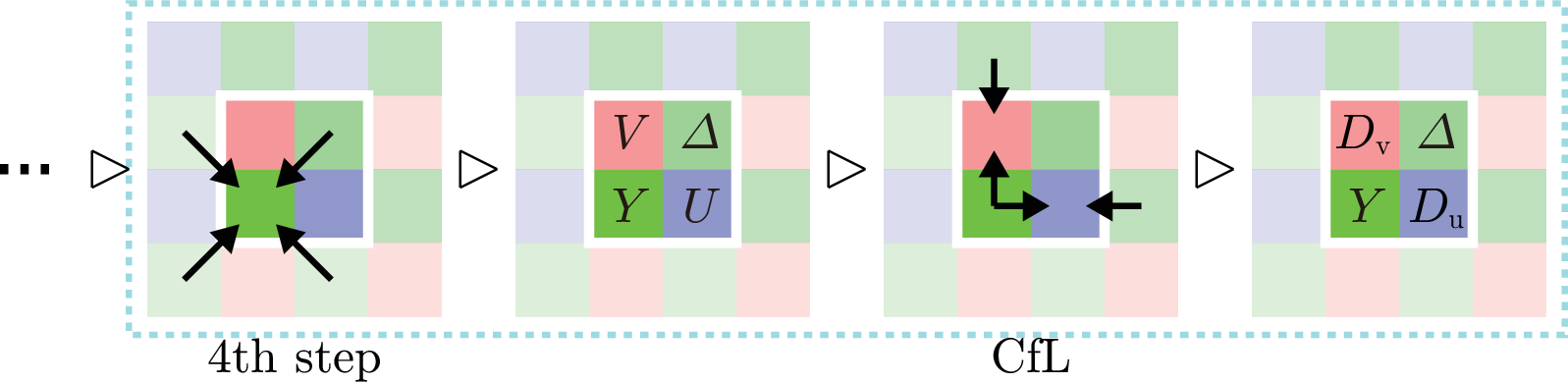}
        \caption{Implementation of CfL-STT (arrows indicate the spatially-aligned CfL operation).}
		\label{Fig_Implement_CfL}
	\end{figure}

To improve raw image compression performance, CfL prediction is integrated into the STT to predict chroma components $U$ and $V$ from the luma component $Y$, resulting in a Y$\Delta$DuDv color space representation (see Fig.~\ref{Fig_Implement_CfL}).
However, as shown in Section~\ref{Sec_Exp}, a direct application of conventional CfL prediction yields image-dependent coding behavior for raw images.
This behavior arises from the lack of spatial alignment between luma and chroma samples and from the sensitivity of least-squares regression to sensor noise, as discussed in Section~\ref{Sec_CfL}.

We introduce a spatially-aligned CfL prediction model defined as
    \begin{equation}
        \widetilde{C}^\star = \a^\star_C~Y_C + \b^\star_C,
    \end{equation}
where $Y_C$ denotes the luma signal spatially aligned with the chroma sampling grid corresponding to component $C$.
The standard CfL prediction formulas in \eqref{Eqs_Alpha_C} and \eqref{Eqs_Beta_C} are modified by replacing $Y$ with $Y_C$, yielding
    \begin{align}
        \a^\star_C
        &=
        \frac{
            N \sum_{n}Y_C(n) C(n)
            -
            \sum_{n}Y_C(n) \sum_{n}C(n)
        }{
            N \sum_{n}Y_C(n)^2
            -
            \left(\sum_{n}Y_C(n)\right)^2
        },
        \label{Eqs_TildeAlpha_C}
        \\
        \b^\star_C
        &=
        \frac{\sum_{n}C(n) - \a^\star_C \sum_{n}Y_C(n)}{N}.
        \label{Eqs_TildeBeta_C}
    \end{align}
When linear interpolation is adopted, the spatially-aligned luma signals are defined as
    \begin{equation}
        Y_U = \frac{Y^l + Y^r}{2}
        \quad\text{and}\quad
        Y_V = \frac{Y^t + Y^b}{2}.
        \label{Eqs_SA_YuYv}
    \end{equation}
These definitions align the luma samples with the chroma sampling grid along the horizontal and vertical directions, respectively.
The interpolation in \eqref{Eqs_SA_YuYv} is not intended to be optimal, but provides a low-complexity and robust spatially aligned luma predictor because the chroma components after STT are located between neighboring luma samples.
It requires only one addition and one division-by-two operation per sample and introduces mild noise suppression without increasing the spatial support.
More elaborate alignment strategies may further improve prediction accuracy, but they would increase computational cost, memory access, and line-buffer requirements.

After parameter estimation, the chroma residuals are defined by
    \begin{equation}
        \Dc^\star = C - \lfloor \widetilde{C}^\star \rfloor.
    \end{equation}
The regression parameters $\alpha_C^\star$ and $\beta_C^\star$ are treated as side information for reconstruction at the decoder.
This formulation preserves integer reversibility for lossless raw image compression.

In the main experiments, $N$ in \eqref{Eqs_TildeAlpha_C} and \eqref{Eqs_TildeBeta_C} denotes all samples in a frame, and the four regression parameters are included as frame-level side information.
This setting evaluates the decorrelation capability of the proposed prediction, but it is not a strict line-based low-latency operation.
Since \eqref{Eqs_TildeAlpha_C} and \eqref{Eqs_TildeBeta_C} can be evaluated over an arbitrary set of samples, the summations can be restricted to samples within a precinct or line group, while the sample prediction in \eqref{Eqs_SA_YuYv} only requires adjacent luma samples.
For clarity, in the experimental evaluation, we refer to the proposed spatially aligned prediction as CfL-STT (Aligned) to distinguish it from a direct application of conventional CfL prediction without spatial alignment.

\section{Experimental Results}\label{Sec_Exp}
To evaluate the proposed CfL-STT (Aligned) in lossless raw image compression, we integrated the CfL-STT module into the JPEG~XS reference software~\cite{JPEGXS_RS} and compared its performance with that of the baseline STT defined in JPEG~XS 2nd Edition~\cite{Descampe2021ProcIEEE}.
Since the original JPEG~XS reference software does not support STT-based lossless compression, we extended it to enable a fully reversible STT-based coding pipeline.
In this pipeline, the non-linear transform (NLT) was disabled to ensure reversibility.
The proposed CfL-STT (Aligned) replaces the standard STT by integrating CfL-based chroma prediction into the transform, resulting in a Y$\Delta$DuDv representation instead of the conventional Y$\Delta$UV.
In addition, a reference method, denoted as CfL-STT (Direct), was evaluated, corresponding to a direct application of conventional CfL prediction without spatial alignment.
In this configuration, CfL prediction is applied to the STT output using luma samples available within each macropixel, which is equivalent to nearest-neighbor interpolation, i.e., $Y_U = Y^l$ and $Y_V = Y^b$.
The test set consisted of ten DNG images from the MIT-Adobe FiveK dataset~\cite{Bychkovsky2011CVPR} and ten DNG images captured using an iPhone SE3~\cite{Suzuki2022TIP}, with bit depths of $12$ or $14$ bits per sample.
Coding efficiency was evaluated in terms of bitrate~[bits per pixel (bpp)] relative to the original CFA-sampled raw image.

\subsection{Decorrelation of Chroma Components}
\begin{figure}[t]
		\centering
        \includegraphics[scale=0.2,keepaspectratio=true]{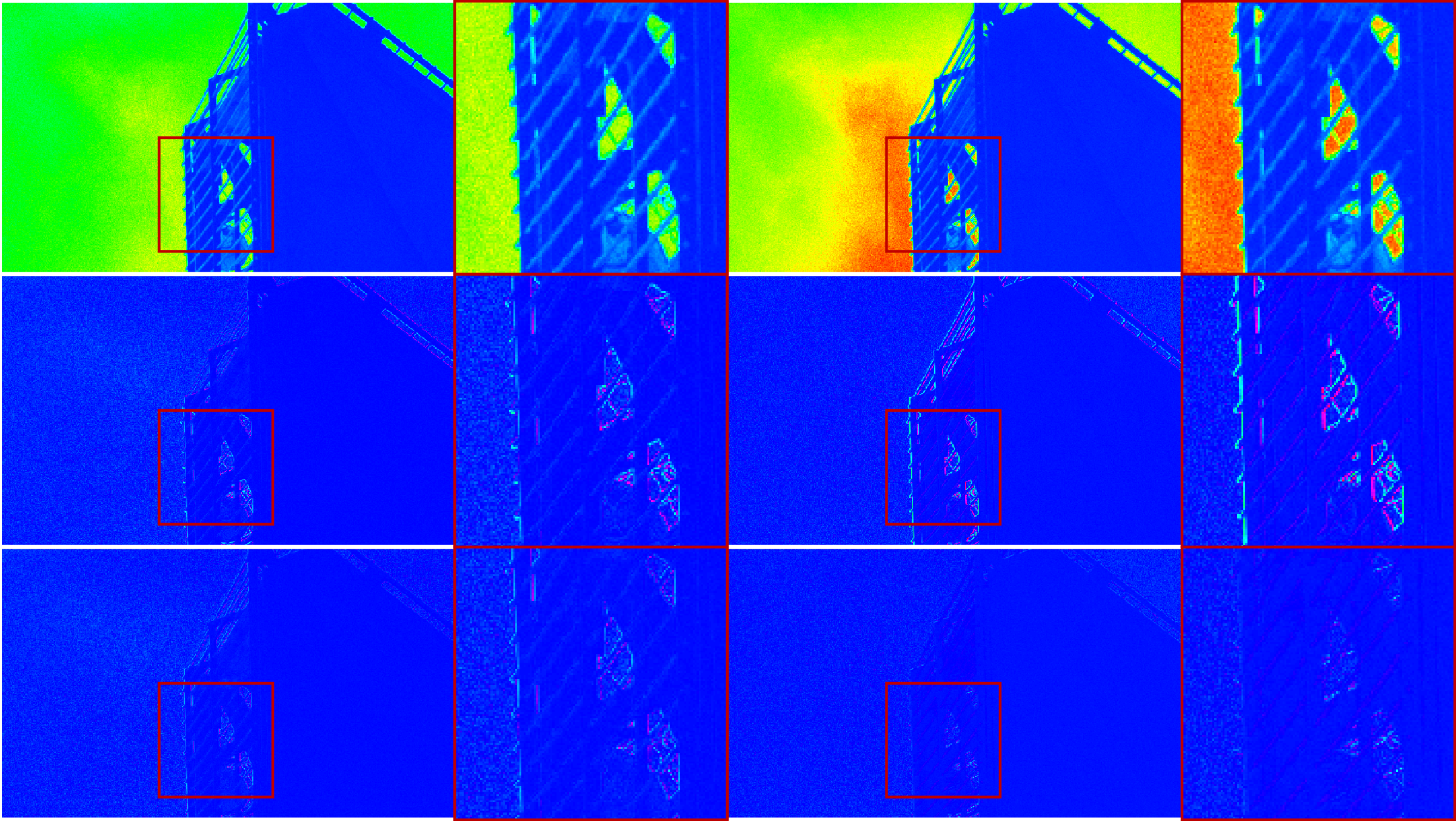}
		\caption{Pseudo-color visualization of a particular area of the decorrelated components for image \textit{m0005} (the same pseudo-color scale and display range are used for all panels, and  colors closer to blue indicate values closer to zero): (top to bottom) $(U,V)$, $(\Du,\Dv)$, and $(\Du^\star,\Dv^\star)$.}
		\label{Fig_Decorr}
	\end{figure}

Figure~\ref{Fig_Decorr} visualizes the decorrelated components produced by STT and the two CfL-STT variants for image \textit{m0005}.
While STT produces chroma components $U$ and $V$, CfL-STT (Direct) yields chroma residuals $\Du$ and $\Dv$, and CfL-STT (Aligned) yields chroma residuals $\Du^\star$ and $\Dv^\star$.

CfL-STT (Direct) exhibited image-dependent coding behavior.
This means that the direct method reduces the bitrate for some images but increases it for others compared with the baseline STT, as quantitatively shown in Section~\ref{Sec_LosslessCompression}.
This image-dependent behavior is caused by the spatial mismatch between luma and chroma samples.
When the nearest luma sample within a macropixel is directly used for CfL prediction, local edges and high-frequency sensor noise are not properly aligned with the chroma sampling positions.
Consequently, the regression model may produce less accurate chroma predictions, broaden the residual distributions, and increase the residual entropy in some cases.

In contrast, CfL-STT (Aligned) uses interpolated luma samples that are spatially aligned with the chroma sampling grids.
This alignment reduces the mismatch between the predictor and the target chroma component while mildly suppressing high-frequency sensor noise.
As a result, CfL-STT (Aligned) produced more concentrated residual distributions, indicating improved decorrelation between luma and chroma components.

\subsection{Lossless Compression}\label{Sec_LosslessCompression}

Table~\ref{Tab_Lossless} reports the lossless compression results.
The overhead of the four frame-level regression parameters is included in the reported bitrates, but it is not shown separately because its effect is negligible at the reported precision.
CfL-STT (Direct) exhibited image-dependent coding behavior and degraded compression efficiency for some images.
In contrast, CfL-STT (Aligned) consistently outperformed the baseline STT for all test images, reducing the average bitrate from $7.00$ to $6.94$~bpp and from $7.29$ to $7.16$~bpp for the two datasets, respectively.
These results indicate that spatial alignment is essential for effectively applying CfL prediction in CFA-domain spectral-spatial transforms.
Although the average bitrate reductions of $0.06$ and $0.13$~bpp are modest, the improvements were observed for all tested images.

We also conducted an encoder-side line-group simulation in which the regression parameters were estimated every $16$ component lines, corresponding to $32$ input raw lines.
Assuming four $32$-bit floating-point coefficients per line group, the side-information overhead was about $10^{-3}$~bpp.
Including this overhead, the average bitrates were $6.93$ and $7.13$~bpp for the two datasets, respectively.
This suggests that the proposed formulation can retain its coding gain under local coefficient estimation without requiring full-frame statistics.

Regarding computational complexity, the additional processing mainly consists of two-tap interpolation, scalar linear prediction, and encoder-side accumulation of simple sums for coefficient estimation.
Since the regression parameters are transmitted as side information, no parameter estimation is required at the decoder.
The additional local memory access is limited to adjacent luma samples, although its interaction with line buffering and the JPEG XS rate allocator is left for future work.
The proposed method therefore provides a practical trade-off between limited additional complexity and bitrate reduction.

Finally, a detailed analysis of backward-adaptive coefficient estimation, lossy coding including quantization effects and bitrate-dependent behavior, and their interaction with the JPEG XS rate allocator is left for future work.

    \begin{table}[t]
    	\centering
    	\caption{Bitrates~[bpp] for lossless compression (bold fonts indicate the best results and blue/red ones indicate improved/degraded results compared to STT).}
        \setlength{\tabcolsep}{4pt}
    	\begin{tabular}{c|ccc||c|ccc}
    	\thline
    	Test & STT & \multicolumn{2}{c||}{CfL-STT} & Test & STT & \multicolumn{2}{c}{CfL-STT} \\
    	Images & \cite{Richter2021TIP} & Direct & Aligned & Images & \cite{Richter2021TIP} & Direct & Aligned \\
    	\hline
    	\textit{\#0500} & $\tn{6.70}$ & $\tr{6.73}$ & $\TB{6.67}$ & \textit{m0001} & $\tn{7.67}$ & $\tr{7.75}$ & $\TB{7.52}$  \\
    	\textit{\#1000} & $\tn{6.70}$ & $\tr{6.72}$ & $\TB{6.67}$ & \textit{m0002} & $\tn{7.57}$ & $\tr{7.68}$ & $\TB{7.43}$  \\
        \textit{\#1500} & $\tn{8.49}$ & $\tb{8.46}$ & $\TB{8.38}$ & \textit{m0003} & $\tn{6.89}$ & $\TB{6.88}$ & $\TB{6.88}$  \\
    	\textit{\#2000} & $\tn{5.51}$ & $\tr{5.56}$ & $\TB{5.45}$ & \textit{m0004} & $\tn{7.21}$ & $\tb{7.20}$ & $\TB{7.08}$  \\
    	\textit{\#2500} & $\tn{6.49}$ & $\tr{6.50}$ & $\TB{6.45}$ & \textit{m0005} & $\tn{6.55}$ & $\tr{6.56}$ & $\TB{6.42}$  \\
    	\textit{\#3000} & $\tn{6.11}$ & $\tr{6.13}$ & $\TB{6.04}$ & \textit{m0006} & $\tn{7.32}$ & $\tb{7.30}$ & $\TB{7.29}$  \\
    	\textit{\#3500} & $\tn{7.68}$ & $\tb{7.65}$ & $\TB{7.57}$ & \textit{m0007} & $\tn{6.62}$ & $\tr{6.65}$ & $\TB{6.44}$  \\
    	\textit{\#4000} & $\tn{8.88}$ & $\tn{8.88}$ & $\TB{8.87}$ & \textit{m0008} & $\tn{8.31}$ & $\tr{8.39}$ & $\TB{8.08}$  \\
    	\textit{\#4500} & $\tn{7.03}$ & $\tr{7.09}$ & $\TB{6.98}$ & \textit{m0009} & $\tn{6.92}$ & $\tb{6.88}$ & $\TB{6.81}$  \\
    	\textit{\#5000} & $\tn{6.36}$ & $\tr{6.40}$ & $\TB{6.27}$ & \textit{m0010} & $\tn{7.80}$ & $\tr{7.88}$ & $\TB{7.61}$  \\
    	\rowcolor[gray]{0.8}
    	Avg.            & $7.00$ & $\tr{7.01}$ & $\TB{6.94}$ & Avg. & $7.29$ & $\tr{7.32}$ & $\TB{7.16}$ \\
        \thline
    	\end{tabular}
    	\label{Tab_Lossless}
    \end{table}

\section{Conclusion}
This study proposed CfL-STT for lossless JPEG XS raw image compression.
The proposed method predicts chroma components from spatially aligned luma samples, producing a more decorrelated Y$\Delta$DuDv representation while preserving exact reversibility.
Experimental results showed that direct CfL prediction can yield image-dependent performance and may degrade coding efficiency, whereas the proposed aligned prediction consistently improves lossless coding efficiency.
The encoder-side line-group simulation further suggests that the coding gain can be retained under local coefficient estimation.
Future work includes complete bitstream-level signaling, interaction with the JPEG XS rate allocator, and more adaptive spatial alignment strategies.

\section*{Acknowledgment}
The authors would like to express their gratitude to C. Huang and K. Oka of our laboratory for their invaluable assistance in the initial implementation of the coding framework used in this study.

\newpage
\bibliography{references} 
\bibliographystyle{IEEEbib} 

\vfill

\end{document}